\documentclass[prb,reprint,superscriptaddress,a4paper,aps,showpacs]{revtex4-1}
\usepackage{graphicx}
\usepackage{tabularx}
\usepackage{amsmath, amsthm, amssymb}

\begin{document}

\newcommand{\m}{$\mu_{\rm B}$}

\title{Evidence for cooper pair diffraction on the vortex lattice of superconducting niobium}

\author{A. Maisuradze}
\affiliation{Physik-Institut der Universit\"at Z\"urich, Winterthuerstrasse 190, CH-8057 Z\"urich, Switzerland}
\affiliation{Laboratory for Muon-Spin Spectroscopy, Paul Scherrer Institute, 5232 Villigen-PSI, Switzerland}

\author {A.~Yaouanc}
\affiliation{Laboratory for Muon-Spin Spectroscopy, Paul Scherrer Institute, 5232 Villigen-PSI, Switzerland}
\affiliation{Institut Nanosciences et Cryog\'enie, SPSMS, CEA and University Joseph Fourier, F-38054 Grenoble, France}

\author{R.~Khasanov}
\affiliation{Laboratory for Muon-Spin Spectroscopy, Paul Scherrer Institute, 5232 Villigen-PSI, Switzerland}

\author{A.~Amato}
\affiliation{Laboratory for Muon-Spin Spectroscopy, Paul Scherrer Institute, 5232 Villigen-PSI, Switzerland}

\author{C.~Baines}
\affiliation{Laboratory for Muon-Spin Spectroscopy, Paul Scherrer Institute, 5232 Villigen-PSI, Switzerland}

\author{D.~Herlach}
\affiliation{Laboratory for Muon-Spin Spectroscopy, Paul Scherrer Institute, 5232 Villigen-PSI, Switzerland}

\author{R.~Henes}
\affiliation{Max-Planck-Institut für Metallforschung, Heisenbergstr. 3, D-70569 Stuttgart, Germany}

\author{P.~Keppler}
\affiliation{Max-Planck-Institut für Metallforschung, Heisenbergstr. 3, D-70569 Stuttgart, Germany}

\author{H.~Keller}
\affiliation{Physik-Institut der Universit\"at Z\"urich, Winterthuerstrasse 190, CH-8057 Z\"urich, Switzerland}

\date{\today}

\begin{abstract}
We investigated the Abrikosov vortex lattice (VL) of a pure Niobium single crystal
with the muon spin rotation ($\mu$SR) technique.
Analysis of the $\mu$SR data in the framework of the BCS-Gor'kov theory allowed us
to determine microscopic parameters and the limitations of the theory.
With decreasing temperature the field variation around the vortex cores deviates
substantially from the predictions of the Ginzburg-Landau theory and adopts a
pronounced conical shape. This is evidence of partial diffraction of Cooper pairs
on the VL predicted by Delrieu for clean superconductors.

\end{abstract}


\maketitle

The Ginzburg-Landau (GL) theory for superconductors is expressed in terms
of an order parameter
$\Delta({\bf r})$. While the absolute value of $\Delta({\bf r})$
determines the local superfluid density, its phase
gradient is proportional to the local magnetic vector potential.
This phase variation leads to the magnetic flux quantization
and the formation of a periodic vortex lattice (VL) in type-II
superconductors as was predicted by Abrikosov.\cite{Abrikosov57}
The VL field variation is uniquely characterized by two length scales,
the magnetic penetration
depth $\lambda$ and the coherence length $\xi_{\rm GL}$.
This simple model turned out to be quite successful in describing
the behavior of a superconductor in a magnetic field\cite{Tinkham96,Ketterson99}
and serves as a basis for data analysis of experiments.\cite{Brandt97,Sonier00,Sonier07}
However, as was shown theoretically
by Delrieu,\cite{Delrieu72} 
the GL model is unable to describe the magnetic response in clean superconductors at  low
temperatures and close to the upper critical field $B_{\rm c2}$.

Soon after the publication of the  microscopic Bardeen-Cooper-Schrieffer (BCS) theory
for conventional superconductors,\cite{Bardeen57} using a Green function
formalism Gor'kov derived the GL equations from the BCS theory.\cite{Gorkov59}
Based on Gor'kov's equation Delrieu analyzed the field variation for
classical $s-$wave superconductors in the vicinity of $B_{\rm c2}$.\cite{Delrieu72}
He found that for clean superconductors the Cooper pairs (CPs) with balistic trajectories
through the vortex cores diffract on the periodic potential of the VL.
As a result, in the low temperature limit close to $B_{\rm c2}$ the spatial field variation
around a vortex core has a conical shape, rather than the cosine-like GL behaviour,
and the fields at the minimum and saddle
points are interexchanged relative to the GL prediction.
Nearly at the same time Brandt came to the same conclusion based on
a nonlocal theory of superconductivity.\cite{Brandt74b467}
To observe the effect of diffraction of CPs,
the carrier mean-free path $\ell_{\rm mfp}$ should exceed the intervortex distance,
the measurements should be done in the vicinity of $B_{\rm c2}(T)$
such that the $\Delta({\bf r})$ gradient is negligible,
and the temperature should be low to minimize thermal fluctuations.

Although the theoretical study of the influence of the diffraction of CPs on the field
variation was already performed in 1972, it was not investigated
experimentally in detail so far.
Early muon spin rotation ($\mu$SR) experiments revealed a linear
high-field tail in the magnetic field distribution $D^{\rm exp}_{\rm c}(B^Z)$,
in agreement with the theoretical expectation.\cite{Herlach90}
However, as noticed recently,\cite{MaisuradzeYaouanc13arXiv} it occurred at an unexpectedly high
temperature. As we note below the temperature stability is crucial in order to
minimize experimental artifacts also leading to a linear high-field tail
in $D^{\rm exp}_{\rm c}(B^Z)$.
On the other hand, a $\mu$SR study of vanadium did not reveal any deviation
from the GL theory.\cite{Laulajainen06} Thus, superconductivity in the
clean limit is one of the critical conditions for the observation of the high-field linear tail.
Most of the novel high temperature superconductors (HTSs) are in the clean limit, and
the tail should be observed provided
the measurements are performed close to $B_{\rm c2}$ and at low temperature.
Such studies of HTSs still await to be performed.

As a first superconductor to look for the effect of CP diffraction,
we have chosen metallic niobium (Nb), since it is a
simple BCS superconductor and pure single crystals
are available. It is a type~II superconductor
($\kappa \approx 0.8 > 1/\sqrt{2} \simeq 0.7$), and therefore
a VL is expected in the bulk when an external
field ${\bf B}_{\rm ext}$ larger than the lower critical field is applied.
As shown by small angle neutron scattering, for $ {\bf B}_{\rm ext}$
parallel to the crystallographic $\langle 111 \rangle$ direction
the VL exhibits a simple triangular lattice.\cite{Schelten71,Kahn73,Muhlbauer09}

Our Nb sample was a single crystal disk of 13~mm diameter and 2~mm thickness
with the $\langle 111 \rangle$ axis oriented normal to the disk.
The samples studied here and in Ref.~[\onlinecite{Herlach90}] come from the same
batch, so they should be of the same metallurgical quality.
The sample is further characterized when discussing its value of $B_{\rm c2} (0)$
(see supplemental materials: Ref.~[\onlinecite{supplemental}]).

The $\mu$SR experiments were performed at the Swiss Muon Source
(S$\mu$S), Paul Scherrer Institute (PSI), Switzerland, using
the general purpose spectrometer (GPS) for $T \geq 1.6$~K and
the low temperature facility (LTF) for $T \leq 1.6$~K.
A field cooled procedure was used with ${\bf B}_{\rm ext}$
perpendicular to the sample plane  (parallel to the
$\langle 111 \rangle$ axis). The $\mu$SR spectra were recorded in
the transverse field geometry, i.e. the initial muon spin polarization
${\bf S}_\mu$ was perpendicular to ${\bf B}_{\rm ext}$. By definition,
${\bf B}_{\rm ext}$ is parallel to the $Z$ axis of the laboratory
orthogonal reference frame. With this geometry the field distribution
in the bulk of a superconductor can be probed $D^{\rm exp}_{\rm c}(B^Z)$.\cite{Yaouanc11}
We explicitly distinguish $D_{\rm c}(B^Z)$ from $D^{\rm exp}_{\rm c}(B^Z)$,
since $D_{\rm c}(B^Z)$ only stands for the distribution of
a perfect VL without crystal disorder.

\begin{figure}
\includegraphics[width=0.99\linewidth,trim=0.6cm 0.6cm 0.8cm 0.8cm, clip=true]{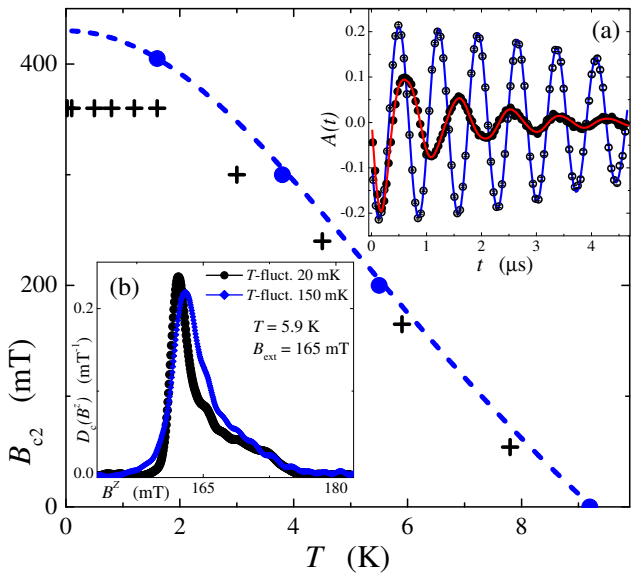}
\caption{(Color online)
$B_{\rm c2}(T) $ for  ${\bf B}_{\rm ext} \parallel \langle 111 \rangle$
as determined by $\mu$SR measurements for our Nb single crystal sample (circles).
The dashed line  corresponds to the equation
$B_{\rm c2}(T) =  B_{\rm c2}(0)(1- \tau^2)/(1 + \tau^2)$ proposed
in Ref.~[\onlinecite{Finnemore66}].
Here $\tau = T/T_{\rm co}$ with $T_{\rm co} = 9.25$~K and
$B_{\rm c2}(0) = 430 \, (2)$~mT. The crosses indicate the points
in the field-temperature diagram at which the field distributions
displayed in Fig.~\ref{fig:distribution_Delrieu} were measured. Insert (a) shows
$\mu$SR asymmetry spectra $A(t)$ recorded at 1.6~K in the normal
($\circ$) and in the mixed ($\bullet$) states for
$B_{\rm ext}$ = 450 and 360~mT, respectively.
The solid lines are fits of Eq.~(\ref{eq:At}) to the data.
The spectra are shown in a rotating frame of 440 and 350~mT,
respectively.
Insert (b) shows the effect of the sample temperature stability on $D^{\rm exp}_{\rm c}(B^Z)$.
The two measurements were performed at $B_{\rm ext}$=165~mT
and $T=5.9$~K with different temperature regulation systems.
For one of them the
temperature was found to oscillate periodically around an average
with a period of a few seconds and a peak to peak amplitude of 150~mK.
For the other system this amplitude was reduced to about 20~mK.}
\label{fig:experimental}
\end{figure}

The forward and backward positron detectors with respect to ${\bf S}_\mu$
were used to build the $\mu$SR asymmetry time spectra $A(t)$
recorded with total statistics ranging from $1.0 \times 10^7$ to
$8.0 \times 10^7$ positron events. Typical $A(t)$ in the normal
and the superconducting states are displayed in the
insert (a) of Fig.~\ref{fig:experimental}.
Note that in contrast to the normal state, a strong damping of $A(t)$ in
the superconducting state is observed which is characteristic of the local magnetic
field variation due to the VL. From these kind of measurements $B_{\rm c2}(T)$
was determined, yielding $B_{\rm c2}(0)=430(2)$~mT (see Fig.~\ref{fig:experimental}).
This value is smaller than  $B_{\rm c2}(0)=443$~mT reported for a sample with a
residual resistivity ratio ${\rm RRR} =  750$.\cite{Williamson70} Hence,
for our sample the ${\rm RRR} > 750$.
Our value of $B_{\rm c2}(0)$ indicates that the sample is of high quality
and pure.\cite{Alekseyevskiy74}

As shown in the insert (b) of Fig.~\ref{fig:experimental} the temperature stability
is important for recording high quality data close to $B_{\rm c2}$.
Large fluctuations of temperature may lead to substantial smearing of the
measured spectra. The  experimental and
theoretical field distributions presented in this letter,
were obtained by Fourier transformation (FT) of the Gaussian
apodized time spectra
(i.e. Fourier transform of $A(t)\exp[-  (t/\sigma_{\rm app})^2/2]$
where $\sigma_{\rm app} = 4.7 \, \mu {\rm s}$). Note that the apodization
has no influence on the analysis, since we directly fit $A(t)$,
rather than $D^{\rm exp}_{\rm c}(B^Z)$.

\begin{figure*}
\includegraphics[width=0.33\linewidth,trim=0.6cm 0.6cm 0.8cm 0.8cm, clip=true]{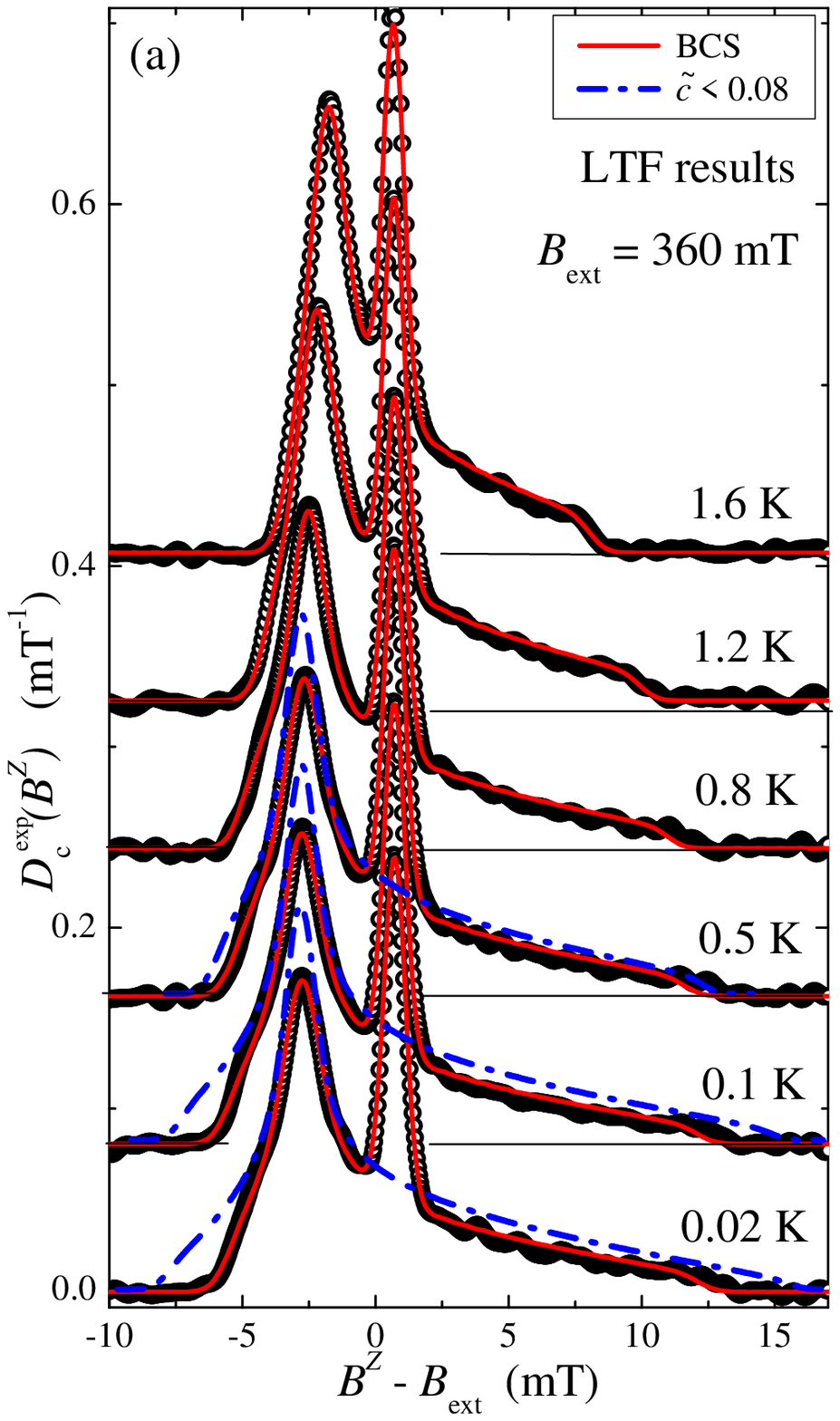}
\includegraphics[width=0.33\linewidth,trim=0.6cm 0.6cm 0.8cm 0.8cm, clip=true]{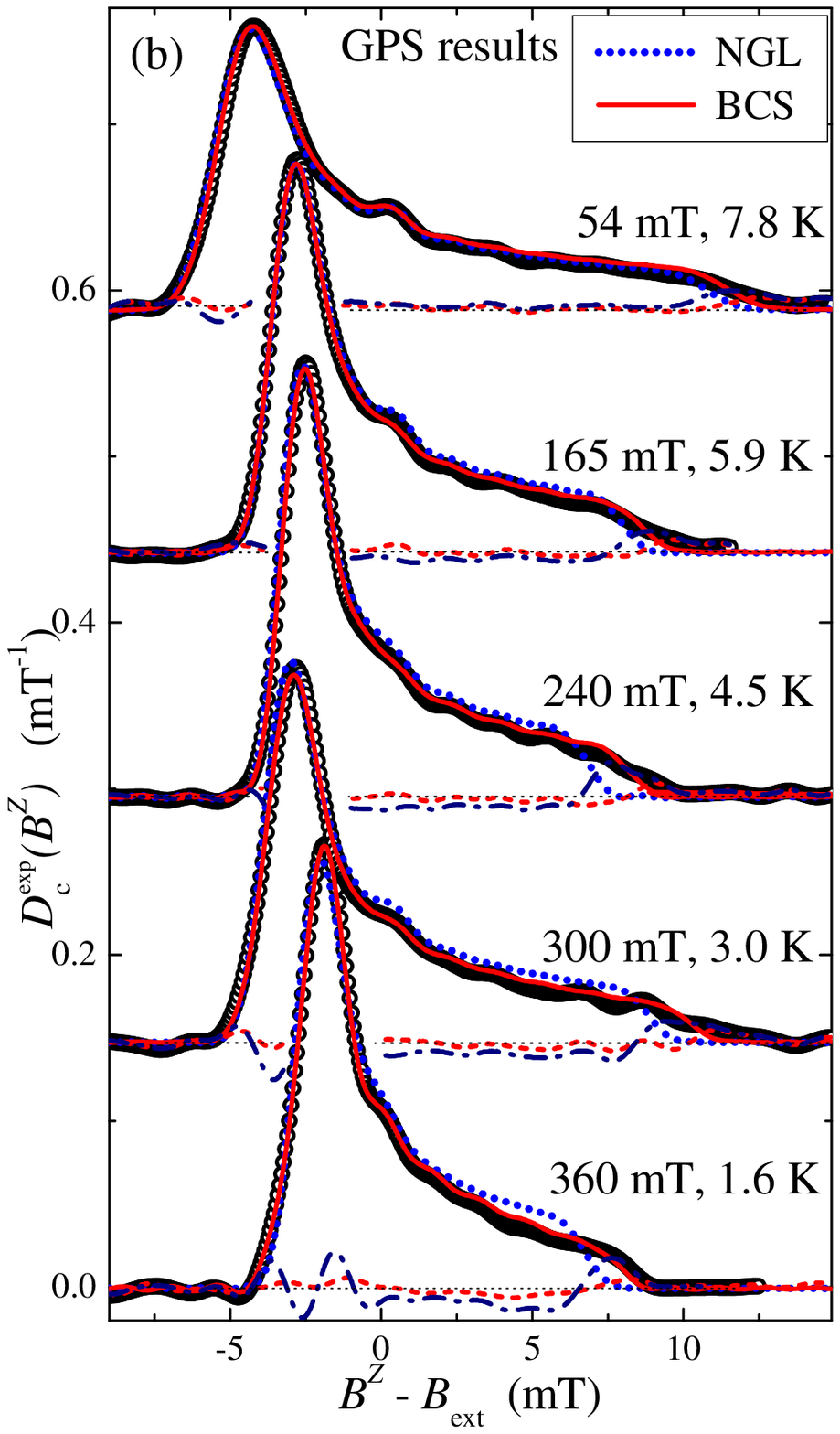}
\includegraphics[width=0.32\linewidth]{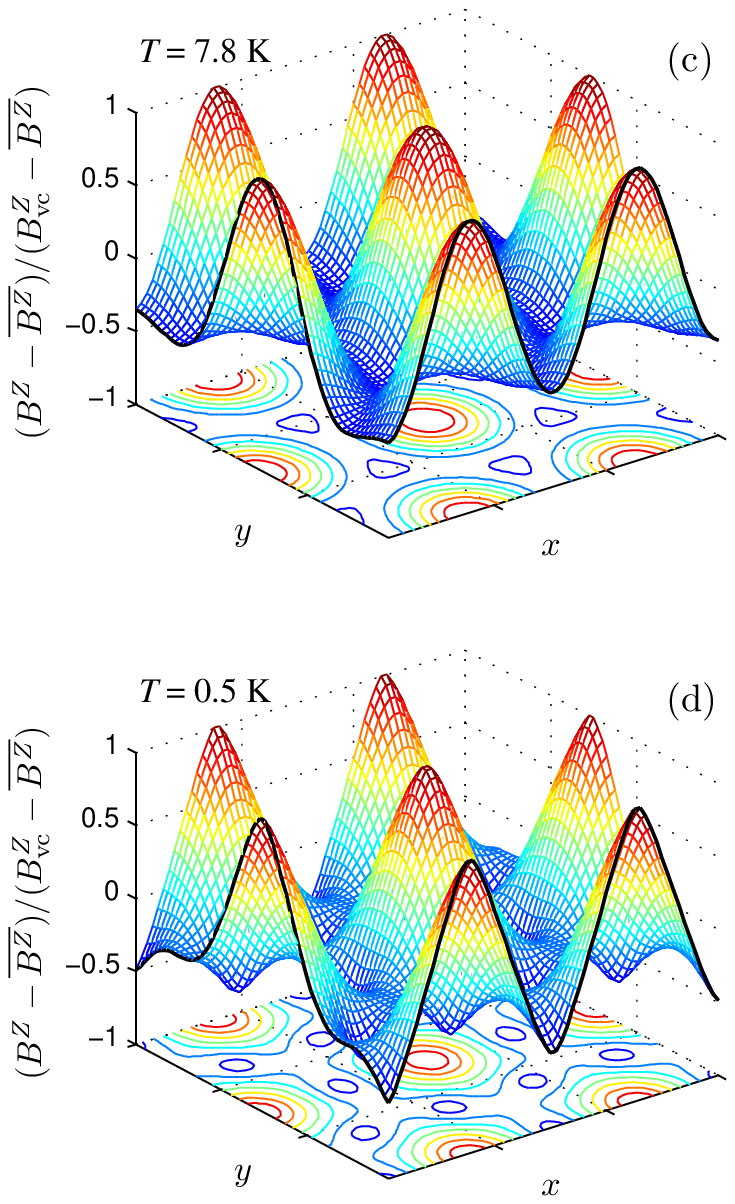}
\caption{(Color online)
Field distributions $D^{\rm exp}_{\rm c}(B^Z)$ of Nb single crystal obtained in the vicinity
of $B_{\rm c2}(T)$ with the LTF spectrometer (a) and the GPS (b).
The solid lines represent  best
fits of Eqs. (\ref{eq:At})-(\ref{Fourier})  to the data.
The fitting parameters are given in the Table~I of Ref.~[\onlinecite{supplemental}].
Leaving the last available parameter $v_{\rm F}$ free,
leads to proper fits for $T \geq 0.8$~K, but not below.
The blue dashed-dotted lines in panel (a) represent the
results for $T= 0.5$, 0.1, and 0.02~K. At 0.5~K the misfits are small,
but still present.
The solid lines for $T \leq 0.5$~K are computed
with the fixed value $\tilde{c}=0.08$ for $T = 0.6$~K.
The blue dotted lines in panel (b) correspond to fits with NGL,
while the red dashed and blue dashed-dotted lines visualize the differences
between $D^{\rm exp}_{\rm c}(B^Z)$ and the BCS-Gor'kov and NGL
predictions, respectively. Panels (c) and (d) illustrate  field variation and
contour plot of $B^Z({\bf r})$ obtained with BCS-Gor'kov model
for the parameters at 7.8 and 0.5~K. $B^Z_{\rm vc}$ denotes the vortex core field.
}
\label{fig:distribution_Delrieu}
\end{figure*}

Figure~\ref{fig:distribution_Delrieu} displays the field distributions
measured in the vicinity of $B_{\rm c2}(T)$  from a
temperature of 7.8 K close to $T_{\rm c0}$ (critical temperature at low field)
down to 0.02~K [see Fig.~\ref{fig:experimental}
for the location of the points (crosses) in the field/temperature diagram].
For each $D^{\rm exp}_{\rm c}(B^Z)$ measured at LTF a relatively
intense sharp peak is present at a field slightly larger
than $B_{\rm ext}$, in contrast to the GPS data for which only a small hump
is found. This field structure (sharp peak and small hump)
originates from the muons stopped in
 the cryostat walls and sample holder (background signal). The
present GPS data are therefore of a much
better quality than previous results  obtained
in the same temperature range showing an intense background signal.\cite{Herlach90}
Qualitatively, a linear high-field tail in $D^{\rm exp}_{\rm c}(B^Z)$ is inferred
at maybe 1.2~K and certainly at the lower
temperatures, but not at higher temperature. On the other hand,
this tail is already seen at 2.6~K in the
published data.\cite{Herlach90,Yaouanc11}

A $\mu$SR spectrum  is described by the sum of two contributions:
\begin{equation}
A(t) =  A_0\cdot [\,F_{\rm s}R_{\rm s}(t) + (1-F_{\rm s})R_{\rm bg}(t)\,],
\label{eq:At}
\end{equation}
where $A_0$ is the total initial asymmetry and $F_{\rm s}$ the
fraction of muons stopped in the sample. From analysis we determine $A_0 = 0.211\, (2)$
[$0.217 \, (2)$] and $F_{\rm s} = 0.993 \, (2)$ [$0.817 \, (2)$]
for measurements carried out with the GPS [LTF] spectrometer.
The function
\begin{equation}
R_{\rm s}(t)  =  e^{-{1\over 2}\sigma_{\rm s}^2 t^2} \int D_{\rm c}(B^Z)
\cos(\gamma_\mu B^Z t + \phi_0) {\rm d} B^Z,
\label{eq:Rt}
\end{equation}
describes the time evolution of ${\bf S}_{\mu}$ in the sample while
$R_{\rm bg}(t)  =  \exp(-\sigma_{\rm bg}^2 t^2/2) \cos(\gamma_\mu B^Z_{\rm bg} t + \phi_0)$
accounts for the background.
Here $\gamma_\mu$ = 851.6~Mrad\,s$^{-1}$\,T$^{-1}$ is the muon
gyromagnetic ratio,  $\sigma_{\rm bg} \simeq 0.22(2) \, \mu{\rm s}^{-1}$
stands for the background damping, and $\phi_0$ is the initial
phase. The mean field for the background
$B^Z_{\rm bg}$ is only slightly different from $B_{\rm ext}$
(see Fig.~\ref{fig:distribution_Delrieu}).
We write  $\sigma^2_{\rm s} = \sigma^2_{\rm nu} + \sigma^2_{\rm dis}$,
where $\sigma_{\rm nu}$ accounts for the damping due to the nuclear $^{93}$Nb spins,
and $\sigma_{\rm dis}$ is the parameter
describing the effect of  VL  disorder. As usual, we assume the effect
of disorder to be modelled by a Gaussian function with a field standard
deviation $\sigma_{\rm dis}/\gamma_\mu$.\cite{Brandt88} Although
this is a crude approximation, we note that the influence of VL disorder
on the high-field tail of $D^{\rm exp}_{\rm c}(B^Z)$ is relatively moderate
compared to its effect on the low-field side.\cite{Maisuradze09}

We determine $D_{\rm c}(B^Z)$ from the real space field map
$B^Z({\bf r})$ of the two-dimensional VL:
$D_{\rm c}(B^Z) = \int_{\rm u.c.} \delta(B^Z({\bf r}) - B^Z) {\rm d}^2 {\bf r}$,
where the integral extends over the VL unit cell.
In terms of its Fourier components,
\begin{eqnarray}
B^Z({\bf r}) & = & \sum_{{\bf K}_{m,h}}
B^Z_{{\bf K}_{m,h}} \exp(i{\bf K}_{m,h} \cdot {\bf r} ),
\label{Fourier}
\end{eqnarray}
where the sum is over the reciprocal space.

First we analyze the data with the numerical solution of the GL (NGL) model
for $B^Z_{{\bf K}_{m,h}}$ using  Brandt's
method.\cite{Brandt97} It only depends on $\lambda$ and
$\xi_{\rm GL}$. The fit for the highest temperature data is reasonable
(see Fig.~\ref{fig:distribution_Delrieu}b). We get $\lambda =59.4\, (2)$~nm
and  $\sigma_{\rm s} = 0.72 \, (1)$~mT, with
$\xi_{\rm GL} = 66.5\, (2)$~nm estimated from the measured
$B_{\rm c2} (7.8~{\rm K})$ and using the GL formula:
$\xi_{\rm GL} = (\Phi_0/2\pi B_{\rm c2})^{1/2}$.
Here, $\Phi_0$ is the flux quantum.
As expected, $\kappa = \lambda/\xi_{\rm GL} = 0.89 \, (1) > 1/\sqrt{2} \simeq 0.7$.
The results for $T \leq 5.9$~K in Fig.~\ref{fig:distribution_Delrieu}b
were obtained with $\lambda$, $\xi_{\rm GL}$, and
$\sigma_{\rm s}$ as free parameters. The GL model fails to describe the
high-field tails in $D^{\rm exp}_{\rm c}(B^{Z})$. In addition, unreasonable
large $\kappa$ values are derived. For example: $\kappa =48.7/28.5= 1.7$ at 1.6~K.
If $\kappa$ had been taken temperature independent as it should, the misfits
are even worse. As expected, the GL model
can only describe $D^{\rm exp}_{\rm c}(B^{Z})$ well near $T_{\rm co}$.

Next we analyze the data with the BCS-Gor'kov theory.\cite{Delrieu72,Delrieu74,MaisuradzeYaouanc13arXiv}
First we discuss the characteristics of $B^Z_{{\bf K}_{m,h}}$  in the
vicinity of $B_{\rm c2}(T)$.
We use the notations of Ref.~[\onlinecite{MaisuradzeYaouanc13arXiv}].
The Fourier component is a function of
four parameters: $B^Z_{{\bf K}_{m,h}} =
f_{m,h}(\tilde{a}, \tilde{b}, \tilde{c},\tilde{d})$.\cite{Delrieu72,Delrieu74,MaisuradzeYaouanc13arXiv}
Here, ${\tilde a} = -\mu_0 N_0 \Delta^2_0 {\tilde c}/2{\overline {B^Z}}$
does not influence the shape of $B^Z({\bf r})$ and therefore $D_{\rm c}(B^Z)$,
but only determines the scale of the field variation. It is proportional to
the density of states at the Fermi level $N_0$ (per spin, volume, and energy),
the quantity $\Delta_0^2 = \overline {| \Delta({\bf r})|^2}$
($\overline {| \Delta({\bf r})|^2}$ is the spacial average of
$ {| \Delta({\bf r})|^2}$), and is inversely proportional
to the  average field $\overline{B^Z({\bf r})} $.
The dimensionless parameters $\tilde{b}$, $\tilde{c}$,
and $\tilde{d}$ determine the  shape of $D_{\rm c}(B^Z)$ and are expressed
by the ratios of four length scales: $\tilde{b} = (\Lambda /\pi\xi^B)^2$,
$\tilde{c}=\Lambda/\xi^T$, and $\tilde{d} = \Lambda/\ell_{\rm mfp}$.
Here, $\Lambda = [\Phi_0/ (2 \pi \overline {B^Z})]^{1/2}$ is a length
parameter proportional to the intervortex distance.
The field and temperature dependent
length scale $\xi^B = \hbar v_{\rm F}/ (\pi \Delta_0)$ diverges at
$\overline{B^Z} \rightarrow B_{\rm c2}$, i.e. $\tilde{b} \rightarrow 0$,
while $\xi^T =\hbar v_{\rm F}/(2 \pi k_{\rm B} T)$. The parameter
$\tilde{c}$ is strongly temperature and field dependent. It vanishes
as $T\rightarrow 0$ and diverges at
$T\rightarrow T_{\rm c0}$.\cite{MaisuradzeYaouanc13arXiv}
Finally, for clean superconductors $\tilde{d}$ is negligibly small, since
$\ell_{\rm mfp}$ significantly exceeds the intervortex distance.

Cooper pair diffraction may influence
$D_{\rm c}(B^Z)$ when three experimental conditions are met:
$\Lambda \ll  \ell_{\rm mfp}$, $\Lambda \ll \pi\xi^B$, and
$\Lambda\ll \xi^T$.\cite{Delrieu72,Delrieu74,MaisuradzeYaouanc13arXiv}
The first condition implies a clean superconductor, the second is
only satisfied in the  vicinity of $B_{\rm c2}(T)$, and the third
one is only possible at low $T$.
Thus, the minimum of \{$\ell_{\rm mfp}$, $\pi\xi^B$, $\xi^T$ \} determines the
effective diffraction length scale of CPs
relative to the intervortex distance $2.693 \times \Lambda$.

The data analysis was done  with $\tilde{b}\simeq 0.110(1-b)/b$ fixed
($b=\overline{B^Z}/B_{\rm c2}(T)\simeq B_{\rm ext}/B_{\rm c2}(T)$).\cite{supplemental,MaisuradzeYaouanc13arXiv}
For Nb we get $0.01 < {\tilde b} < 0.02$, except for the
spectrum at 7.8~K for which ${\tilde b} = 0.04$. Since
\begin{equation}
{\tilde c} = \frac{ \sqrt{\Phi_0 2\pi} k_{\rm B} T}{\sqrt{\overline{B^Z}} \hbar v_{\rm F} }
           \simeq \frac{ \sqrt{\Phi_0 2\pi} k_{\rm B} T}{\sqrt{B_{\rm ext}} \hbar v_{\rm F} },
\label{parametre_b_c}
\end{equation}
the $B^Z_{{\bf K}_{m,h}}$ depends on ${\tilde a}$, $\tilde{d}$, $v_{\rm F}$, $T$,
and $B_{\rm ext}$.\cite{supplemental}

The analysis of $A(t)$ for $T \geq 0.8$~K
shows that $\tilde{d}\lesssim 0.01$, which agrees with the estimate
of $\ell_{\rm mfp} \simeq 7$ $\mu$m for the Nb sample with RRR$\,=750$.\cite{Williamson70}
Consequently, we are in the clean limit and $\tilde{d}$ has a negligible influence on $D_c(B^Z)$.
The results of the analysis are presented
in Fig.~\ref{fig:distribution_Delrieu} and Table~I of Ref.~[\onlinecite{supplemental}].
The BCS-Gor'kov model describes the high-field
tail significantly better at low temperatures while at the highest temperature both
models reproduce the data equally well. The deviation from the GL theory
and the gradual disappearance of the cut-off singularity at the maximal field is
a result of the conical shape of the spatial field variation at the vortex
cores (see Fig.~\ref{fig:distribution_Delrieu}d), which in turn
is a consequence of the partial CP pair diffraction.
This deviation cannot be explained by the presence of
significant temperature and field fluctuations resulting in a large
smearing parameter $\sigma_{\rm s}$.\cite{Brandt88}
If produced artificially the cut-off singularity in $D^{\rm exp}_{\rm c}(B^Z)$ at $B^Z_{\rm vc}$ vanishes
as is the case in the distribution labeled 150 mK in
insert (b) of Fig. \ref{fig:experimental}.\cite{supplemental, Maisuradze09}
Based on the generalized Bloch equations\cite{Torrey56,Seeger79,Slichter96} and
the analysis\cite{Kubo97,Yaouanc11} of zero-field $\mu$SR result we found  that the
muon diffusion\cite{Schwarz86,Wipf87,Grasselino13} is negligible in the
studied Nb sample (see Ref.~[\onlinecite{supplemental}]). A weak pinning\cite{supplemental}
excludes also an influence of the peak effect\cite{Daniilidis07,Hanson11,Qviller12}
on the measured field distributions.
The BCS-Gor'kov model breaks down for $T \leq 0.5$~K as shown by the
dashed-dotted lines in Fig.~\ref{fig:distribution_Delrieu}a.
A proper description requires to
consider $B^Z_{{\bf K}_{m,h}}$ as a function of ${\tilde a}$ and $\tilde{c}$
rather than of ${\tilde a}$ and $v_{\rm F}$, and to keep the value of $\tilde{c}$
at $T = 0.6$~K for the lower temperatures (solid lines in Fig.~\ref{fig:distribution_Delrieu}a).\cite{supplemental}
This means that the sharpness of the $B^Z({\bf r})$ cones is limited
by a physical process. Refering to Eq.~(\ref{Fourier}), we suggest that the
VL structural disorder may round off the cones, as observed experimentally.

We get $v_{\rm F}  = 2.0 \, (2) \times 10^5$~m/s from the fits for $T \geq 0.8$~K.
This value is compatible with $v_{\rm F}  = 2.73 \times 10^5$
and $2.94 \times 10^5$~m/s
determined from magnetization measurements.\cite{Kerchner81,Williamson70}
From the measured ${\tilde a}$ and ${\tilde c}$
we determine the condensation energy
$ E_c = -2\overline{B^Z}\tilde{a}/\mu_0\tilde{c} = 2\times N_0\Delta_0^2/2$.\cite{Tinkham96,Comment}
While $N_0$ is a constant, $\Delta_0^2$ is field and temperature dependent.
At $T=0$ and interpolating $\Delta_0^2$ to zero field with a conventional
formula,\cite{MaisuradzeYaouanc13arXiv}
$[-2\overline{B^Z}  \tilde{a}/\mu_0\tilde{c}]/[1 - (\overline{B^Z}/B_{\rm c2}(0))] = N_0\Delta_0^2(0)$ where
$\Delta_0(0)$ is the s-wave BCS gap which is
temperature independent for  $T\rightarrow 0$.\cite{Tinkham96}
From $\Delta_0(0) = 1.45$ meV (see Ref.~[\onlinecite{Finnemore66}]) and our estimate for the ground
state $E_c(0)=N_0\Delta_0^2(0) = 2.47(9)\times 10^4$~Jm$^{-3}$, we obtain
$N_0 \approx 4.6\times 10^{47}\,\,\, {\rm J}^{-1}{\rm m}^{-3}{\rm spin}^{-1}
\approx 1.3\,\,\,  {\rm eV}^{-1}{\rm atom}^{-1}{\rm spin}^{-1}$.
Considering the approximate nature of the linear field interpolation,
these results are quite close to the specific heat result
$N_0 = 0.85\,\,\, {\rm eV}^{-1}{\rm atom}^{-1}{\rm spin}^{-1}$ (see Ref.~[\onlinecite{Junod86}])
and the GL condensation energy $B^2_{\rm c}/2\mu_0 = 1.6\times 10^4$ Jm$^{-3}$ for
the thermodynamic critical field $B_{\rm c}=0.20$ T.\cite{Finnemore66}
In our measurements all the conditions for the observation of partial CP
diffraction are met at $T\lesssim 1.2$ K:
$\Lambda/\pi\xi^B \leq 0.11\ll 1$, $\Lambda/\ell_{\rm mfp}\lesssim 0.01\ll 1$,
and $\Lambda/\xi^T = 0.16 \ll 1$.

To conclude, we investigated the magnetic field distribution for the
vortex lattice (VL) of a pure Nb single crystal with the $\mu$SR
technique. The data were analyzed using the solution of the
BCS-Gor'kov equation proposed by Delrieu,\cite{Delrieu72} a microscopic description
in contrast to the conventional GL picture. As a result, we found
strong evidence for  partial Cooper pair (CP) diffraction on the periodic potential of
the vortex lattice reflected in the conical narrowing of the real space field
variation around the vortex cores and in the presence of
a high-field linear tail in the field distribution down to 0.02~K, as expected by the
BCS-Gor'kov theory. However, the BCS-Gor'kov  description is only partially successful
as the prediction for the low-field tail at zero-temperature limit deviates
from the experimental observation, presumably due to the residual VL disorder.
From the analysis we determined
the Fermi velocity $v_{\rm F}  = 2.0 \, (2) \times 10^5$~m/s and the
ground state condensation energy  $N_0\Delta_0^2(0) = 2.47(9)\times 10^4$~Jm$^{-3}$ which
are in reasonable agreement with literature results.\cite{Kerchner81,Williamson70,Finnemore66}
The observation of partial CP diffraction
should not be restricted to Nb. Under proper experimental conditions it should also be
seen for any clean type-II superconductor.

This work was performed at the Swiss Muon Source
(S$\mu$S), Paul Scherrer Institut (PSI), Switzerland and partly supported by
NCCR MaNEP sponsored by the Swiss National Science Foundation.

\newpage

\section{Supplemental materials}

{\it {\bf Abstract:} Based on the phenomenological Ginzburg-Landau (GL) model and generalized Bloch equations for a diffusing magnetic
probe we estimate a possible effect of the muon diffusion on the probability field distribution in Nb.
We show that in the present experiments the effect of diffusion is negligibly small and cannot
have any appreciable effect on the TF $\mu$SR results given in the main text. Extremely
low pinning in the studied Nb sample is demonstrated. Additional information on the data analysis is given.}
\vspace{0.3cm}

\subsection{Effect of muon diffusion on ${\rm Nb}$ $\mu$SR spectra}

As reported by Schwarz {\it et al.} and Wipf {\it et al.} in previous studies of Nb the
diffusion of a light interstitial particle such as a positive muon is possible.\cite{Schwarz86,Wipf87}
As a result a muon probes the magnetic field not at a specific site in the sample
but the field is averaged over a diffusion trajectory.
In order to study a possible effect of muon diffusion on TF $\mu$SR experiments presented
in the main text we first determine the muon diffusion coefficient $D_{\mu}$ from zero field
(ZF) $\mu$SR experiment on the Nb sample studied here. Then we simulate the effect of
muon diffusion on TF spectra using generalized Bloch equations for a diffusing magnetic probe.

\begin{figure}
\includegraphics[width=0.97\linewidth,trim=0.5cm 0.5cm 0.5cm 0.5cm, clip=true]{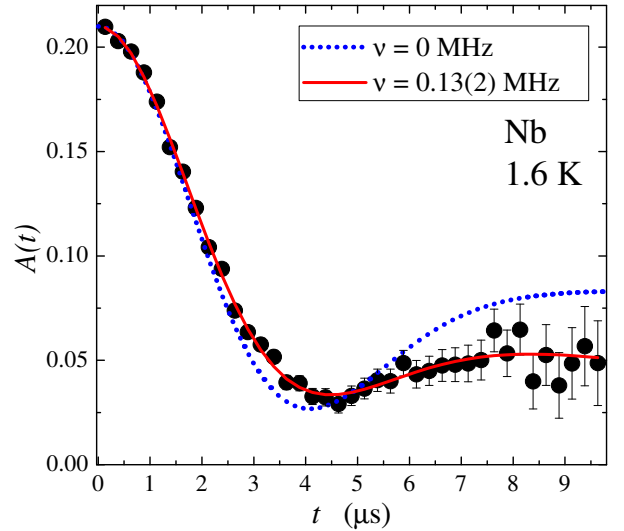}
\caption{(Color online)
Zero field time spectrum at 1.6 K of the Nb single crystal. The solid lines are fit of
the data using Eqs. (\ref{eq:KuboT})-(\ref{eq:ZFasyModel})}
\label{fig:ZF}
\end{figure}

Muon hopping rate of $\nu \sim 0.7$ MHz was determined at 2.5 K
by Grasselino {\it et al.} from ZF $\mu$SR experiments.\cite{Grasselino13}
In zero field the muon experiences a spin depolarization due to
nuclear magnetic fields which for a static case is described with the Kubo-Toyabe function:\cite{Kubo97}
\begin{equation}
P_{\rm KT}(t) = \frac{1}{3} + \frac{2}{3}
\left(1-\sigma^2_{\rm KT}t^2\right)\exp\left(-{\sigma^2_{\rm KT}t^2\over 2}\right)\,.
\label{eq:KuboT}
\end{equation}
The strong collision model was used in Ref.~\onlinecite{Grasselino13} for estimate of the hopping
rate $\nu$. Muon spin polarization within this model is described as follows:\cite{Yaouanc11}
\begin{equation}
P_{\nu,{\rm KT}}(t) =  P_{\rm KT}(t)e^{-\nu t} + \nu \int_0^t P_{\nu,{\rm KT}}(t-t')P_{\rm KT}(t')e^{-\nu t'}dt'\,.
\label{eq:KuboTdyn}
\end{equation}
The result of ZF $\mu$SR measurement at 1.6 K is shown in Fig. \ref{fig:ZF}.
Time dependent asymmetry spectra are analyzed with the expression:
\begin{equation}
A_{\rm ZF}(t) =  A_{\rm S,ZF} P_{\nu,{\rm KT}}(t) + A_{\rm bg,ZF}
\label{eq:ZFasyModel}
\end{equation}
and is shown with the solid line. The model describes the measured spectrum
well. For comparison with dotted line the spectrum for $\nu=0$ is also shown.
Above, the first term describes the signal of the Nb sample while the second term is
the silver background signal. Initial asymmetries are found to be $A_{\rm S,ZF} = 0.190(1)$
and $A_{\rm bg,ZF}  =0.020(1)$. ZF nuclear depolarization rate and muon hopping rate are
$\sigma_{\rm ZF} = 0.424(5)$ $\mu$s$^{-1}$ and $\nu=0.13(2)$ MHz, respectively.
This hopping rate is somewhat smaller than that reported in Ref.~\onlinecite{Grasselino13}.
However, as we can see from  comparison of ZF spectra measured in the present work and that of
Ref.~\onlinecite{Grasselino13}, Fig. 6(a) the relaxation is indeed smaller.

The diffusion coefficient of a particle  in classical statistical physics is
related to its mean velocity  $v = \ell \nu$ and its mean free path $\ell$ through
the following relation:
\begin{equation}
D_{\mu} = \frac{1}{3}v\ell = \frac{1}{3}\nu\ell^2.
\end{equation}
The lattice constant of metallic Nb and consequently the maximal distance between neighboring
muon stopping sites is 0.33 nm. Therefore, for a muon hopping rate of 0.13 MHz and a mean free
path of the order of lattice constant we obtain the diffusion coefficient
$D_{\mu} = 0.0047$~nm$^2$$\mu$s$^{-1}$ at 1.6 K.
This value is two orders of magnitude smaller than the TF $\mu$SR sensitivity
to muon diffusion due to the finite muon lifetime:\cite{Schwarz86} 1~nm$^2\mu$s$^{-1}$
and it is four orders of magnitude
smaller than 80(20)~nm$^2$$\mu$s$^{-1}$ reported by Schwarz {\it et al.}\cite{Schwarz86}
This discrepancy can be explained by number of effects which were not considered in the
analysis of the field distribution smearing around a vortex core field in Ref.~\onlinecite{Schwarz86}.
Namely, at present it is well established that in TF $\mu$SR besides
muon diffusion the smearing of $D_c(B^Z)$  is substantially influenced by a large number
of other factors, such as: gradients and fluctuations of applied
field and temperature, various kinds of vortex disorder, and nuclear fields.\cite{Brandt88}
This led to a substantial overestimate of muon diffusion
rate and consequently $D_{\mu} = 80(20)$~nm$^2$$\mu$s$^{-1}$ (at 4.6~K) can be considered as
the uppermost limit of a possible muon diffusion in Nb.

Next we analyze the effect of a small muon diffusion on TF $\mu$SR field distributions obtained
in the main text.
For a diffusing muon the complex TF muon spin polarization
$\tilde{P}^+_{\rm TF}({\bf r},t)  = \tilde{P}_x({\bf r}, t) + i \tilde{P}_y({\bf r},t)$ can be found
from the solution of the following generalized Bloch equations:\cite{Torrey56,Seeger79}
\begin{equation}
{\partial \tilde{P}^+_{\rm TF}({\bf r},t) \over \partial t} =
-{\tilde{P}^+_{\rm TF}({\bf r}, t) \over T_2} -i\gamma_{\mu}B^Z({\bf r}) + D_{\mu}\nabla^2\tilde{P}^+_{\rm TF}({\bf r}, t),
\label{eq:BlochTF}
\end{equation}
Here, $T_2^{-1}$ is the transverse muon spin relaxation rate.
This rate is negligibly small, and therefore we shall suppress it.
Measured TF muon spin polarization can be found as the  real
part of the spatially averaged complex polarization:
\begin{equation}
P_{\rm TF}(t) = {\mathcal R}e
\left\{ {1\over V_{\rm u.c.}}  \int_{V_{\rm u.c.}} \tilde{P}^+_{\rm TF}({\bf r},t)d^2 {\bf r}\right\}.
\label{eq:Polarization1}
\end{equation}
For a small muon diffusion rate $D_{\mu}$,
when $\gamma_{\mu} D_{\mu} {\partial^2 B^Z \over \partial k^2} t^2 \ll 1$
(here $k = x$, $y$, or $z$) the solution of this differential equation
is:\cite{Slichter96, Seeger79}
\begin{equation}
\tilde{P}_{\rm TF}^+({\bf r}, t) =
 P_{0, {\rm TF}}^+ \exp\left[ 
-{\gamma^2_{\mu}D_{\mu}[\nabla B^Z({\bf r})]^2 t^3 \over 3}
- i\gamma_{\mu }B^Z({\bf r})t  \right].
\label{eq:BlochSolutionTF}
\end{equation}

\begin{figure}
\includegraphics[width=0.45\linewidth,trim=0.6cm 0.06cm 0.8cm 0.8cm, clip=true]{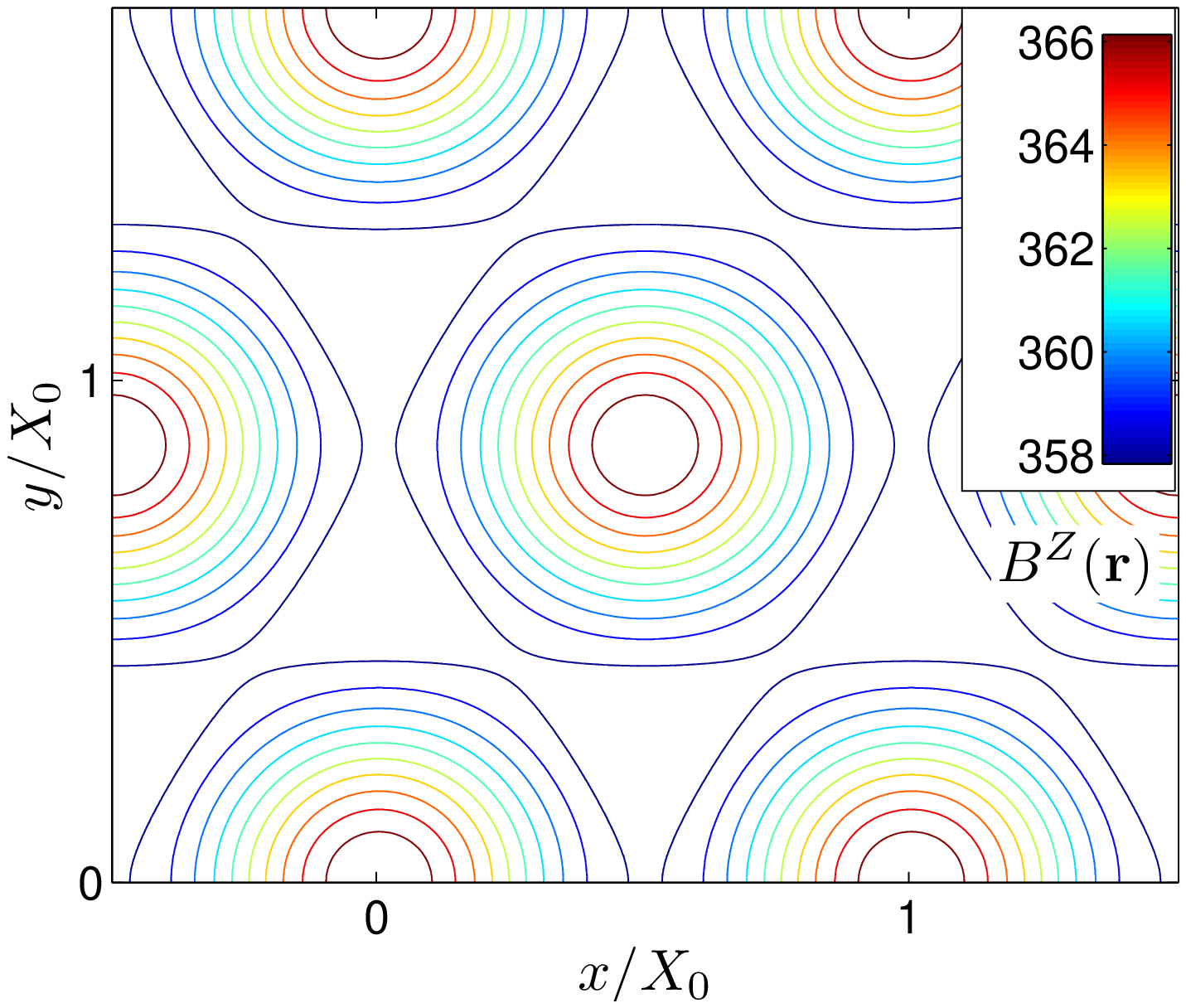}
\includegraphics[width=0.45\linewidth,trim=0.6cm 0.06cm 0.8cm 0.8cm, clip=true]{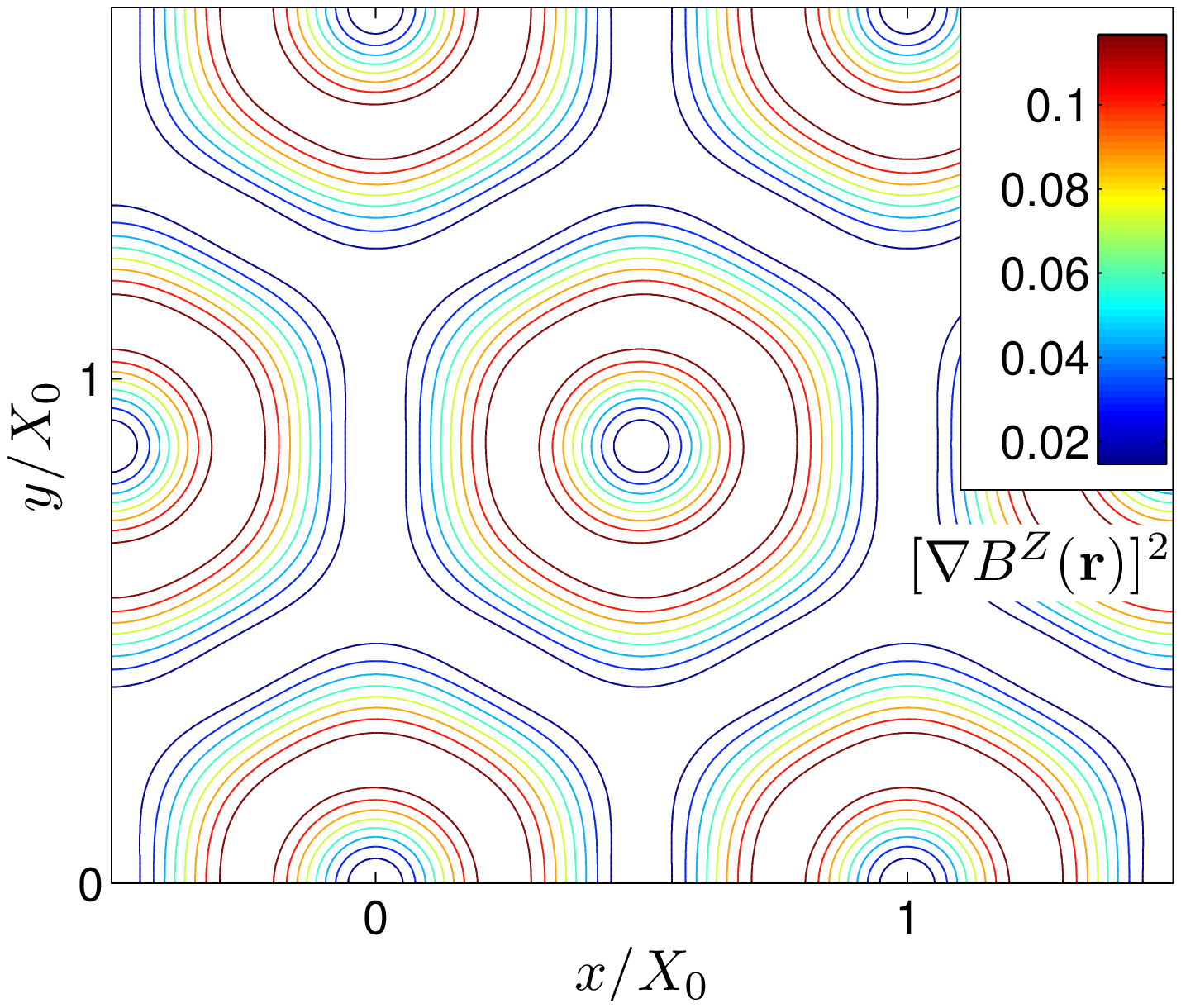}\\
\includegraphics[width=0.45\linewidth,trim=0.6cm 0.06cm 0.8cm 0.8cm, clip=true]{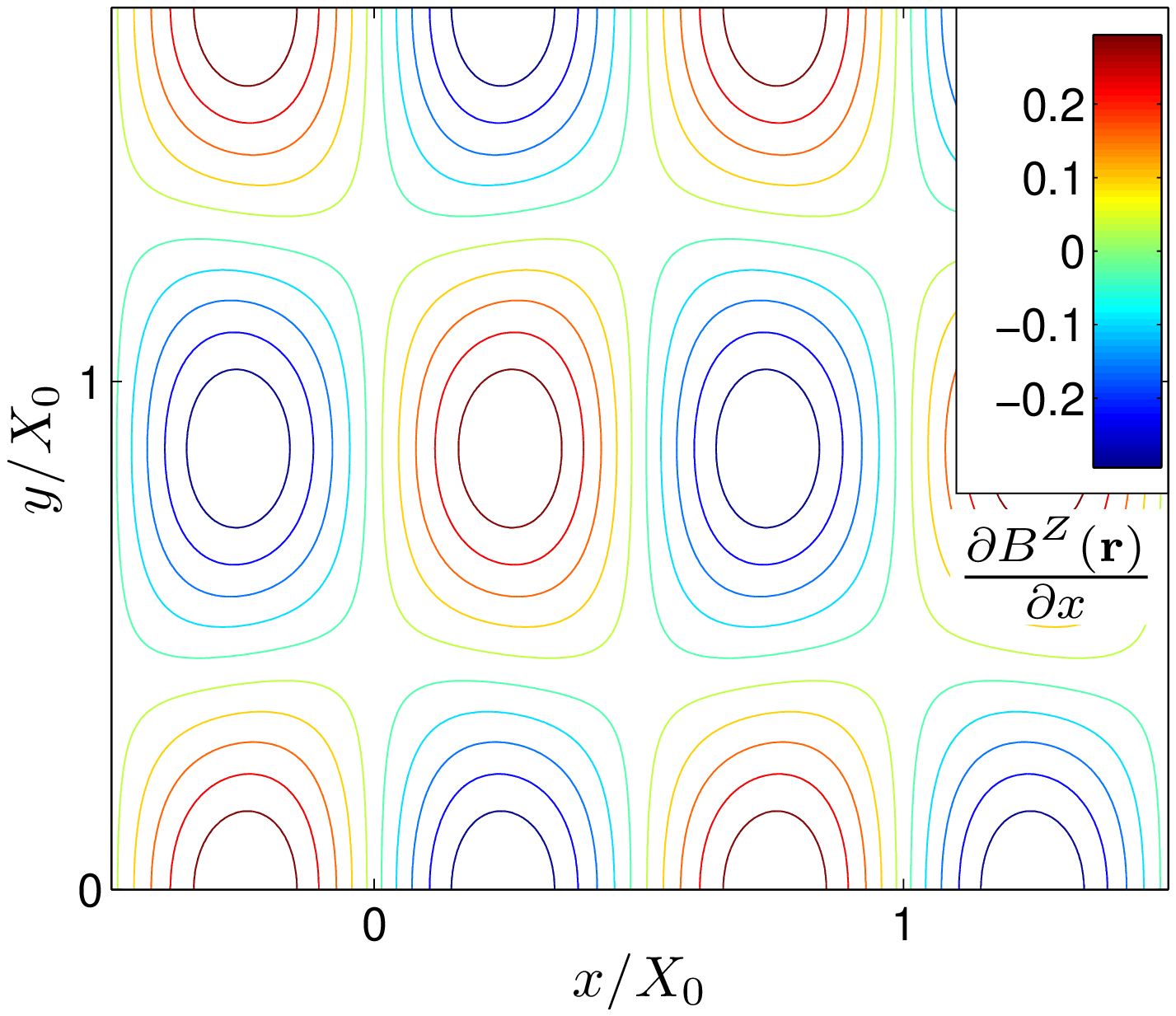}
\includegraphics[width=0.45\linewidth,trim=0.6cm 0.06cm 0.8cm 0.8cm, clip=true]{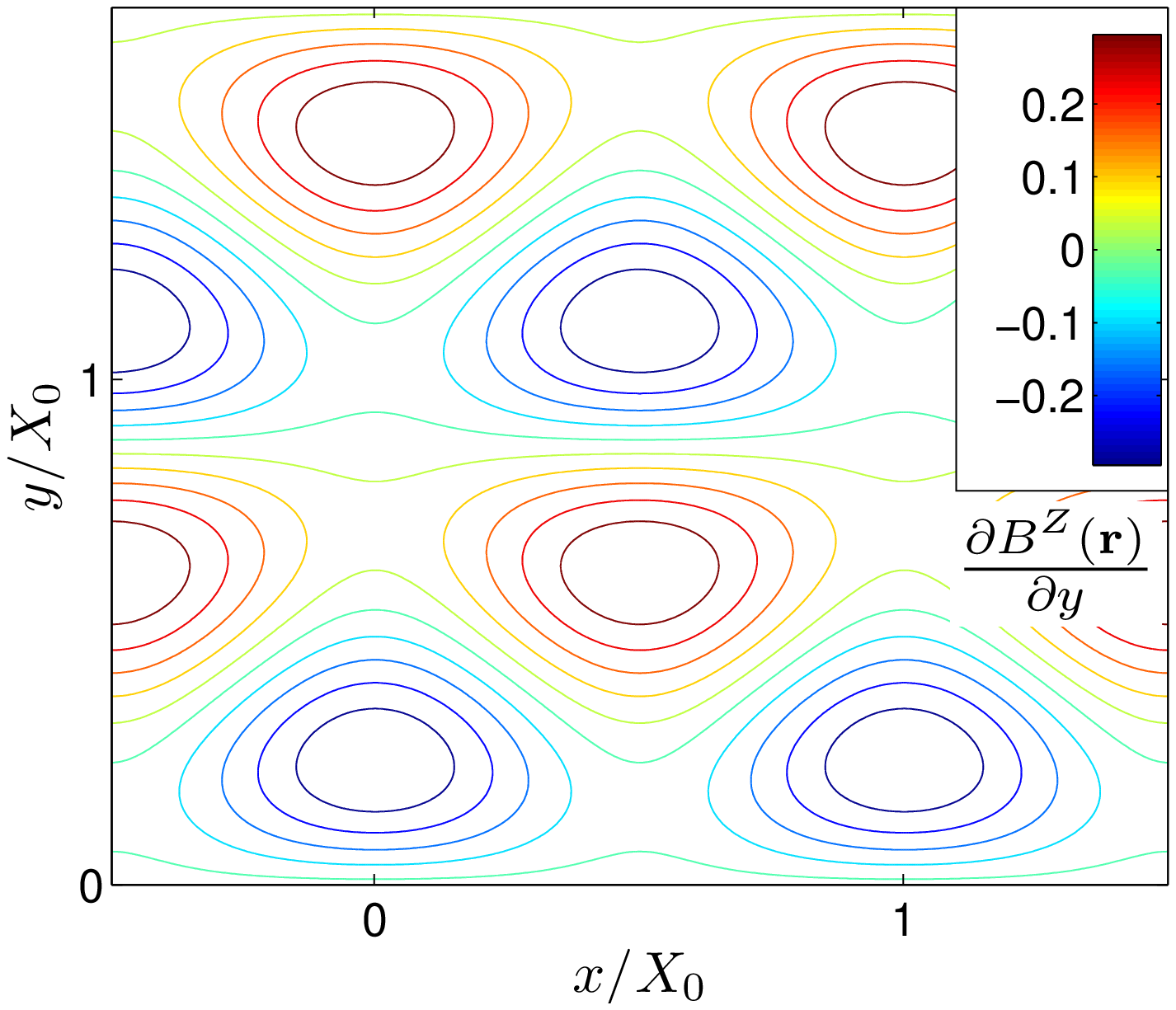}
\caption{(Color online) Contour plot of $B^Z({\bf r})$, $[\nabla B^Z({\bf r})]^2$,
$\partial B^Z({\bf r})/\partial x$,
and $\partial B^Z({\bf r})/\partial y$ obtained using the NGL model for the spectrum at 1.6 K and 360 mT
($\lambda = 48.7$ nm and $\xi = 28.5$ nm). The fields are given in units of mT while corresponding
derivatives are in mT/nm and (mT/nm)$^2$. }
\label{fig:BrDers}
\end{figure}

\begin{figure}
\includegraphics[width=0.9\linewidth,trim=0.6cm 0.06cm 0.8cm 0.8cm, clip=true]{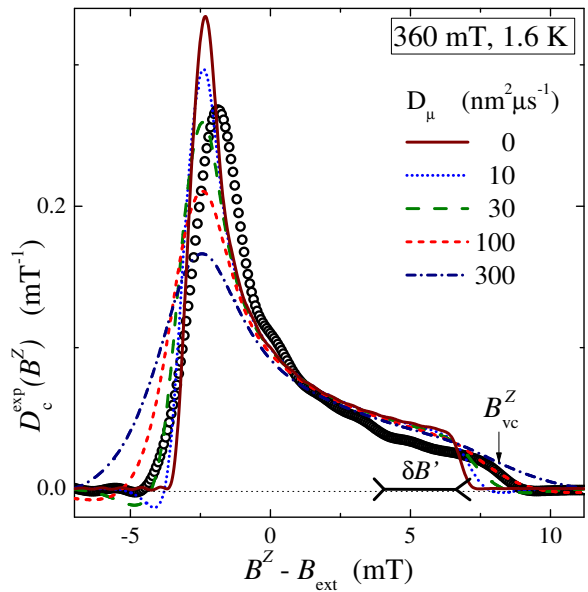}
\caption{(Color online) Effect of the muon diffusion on the field distribution assuming the
NGL model for the vortex field distribution.
Circles illustrate the experimental
spectrum measured at 1.6 K and 360 mT on the GPS spectrometer.
Lines result from the computation of the field distribution with the NGL model
for the vortex field distribution and using Eq.~\ref{eq:BlochSolutionTF} for the
description of the muon diffusion.
The figure shows that increasing the diffusion coefficient $D_\mu$  leads
to a smearing of the cut-off singularities. However, none of the computed
distributions looks like  the measured distribution. }
\label{fig:DcBdiff}
\end{figure}

Next we  show that diffusion smearing cannot lead to a good description
of the spectra if the NGL model is used.
We simulate the influence of the muon diffusion on $D_c(B^Z)$ and compare with our
measured field distribution.
Only the BCS-Gor'kov model can explain the measured
field distribution, in particular the high-field tail.
We proceeded as follows. First the magnetic field and its
gradient was computed [$B^Z({\bf r})$ and $\nabla B^Z({\bf r})$]
using numerical solution of GL for $\lambda = 48.7$~nm and $\xi = 28.5$~nm as obtained for
spectrum measured at 1.6 K and $B_{\rm ext} = 360$ mT in the main text (see Fig.~\ref{fig:BrDers}).
Then the TF muon spin polarization was computed using
Eqs.~(\ref{eq:Polarization1}) and (\ref{eq:BlochSolutionTF}).
Finally, the field distribution was obtained by Fourier
transformation of the apodized polarization function, as explained in
the main text.

The results of the computations are shown in Fig.~\ref{fig:DcBdiff}. In solid line
we show the curve for $D_\mu =0$ nm$^2\mu$s$^{-1}$. As expected the curve
is not a proper description of the measured distribution.
Note that this curve differs from that shown for 1.6 K in the main text,
since here the Gaussian smearing originating from nuclear fields and vortex disorder is
set to zero. Thus, Fig.~\ref{fig:DcBdiff} demonstrates an effect of smearing
solely due to muon diffusion.
With increasing $D_\mu$ smearing of the field distribution increases. But still none
of the computed distribution approaches closely the measured
distribution. Interestingly, a linear high-field tail is predicted at
$D_\mu =3 00$~nm$^2\mu$s$^{-1}$. However, as deduced from our
zero-field measurements, it is not justified to consider large
$D_\mu$ values.

Finally, we would like to stress that important is not only a linear tail but also the spectral
weight of the field distribution close to $B^Z_{\rm vc}$. Muon probes vortex lattice space
randomly. Thus, the probability to find a muon at a distance $r$ from the vortex core is
proportional to the circular area: $P(r){\rm d}r \propto 2\pi r {\rm d}r$. For $T\rightarrow 0$
and $B_{\rm ext}\rightarrow B_{\rm c2}$  the BCS field profile along
the line crossing two neighboring vortex cores  has a sawtooth-like shape while for the NGL
model it is a cosine function (see these profiles e.g. in Ref.~\onlinecite{MaisuradzeYaouanc13arXiv}).
With a good approximation the field in the vicinity of the vortex core depends only on $r$ and can be
expanded in Taylor series: $B^Z(r)=B^Z_{\rm vc}+B'r$ (for BCS) and $B^Z(r)=B^Z_{\rm vc}+0.5B''r^2$
(for NGL). Here, derivatives $B'<0$ and $B''<0$, with $r>0$.
Inverse of these functions are:
\begin{align}
r(B^Z)& = {\delta B^Z \over -B'} &\,\,\,{\rm for~BCS\,\,}\nonumber \\
r(B^Z)& = \sqrt{2\delta B^Z \over -B''}&\,\,\, {\rm for~NGL.} \nonumber
\end{align}
Here we denote $\delta B^Z = B^Z_{\rm vc}-B^Z(r)$.
Consequently the probability field distributions in the vicinity of a vortex core can be
approximated as follows:
\begin{align}
D^{\rm BCS}_{\rm c}(B^Z){\rm d}B^Z\propto-2\pi r{\rm d}r& = 2\pi { \delta B^Z\over B'^2}{\rm d}B^Z &{\rm for~BCS\,\,\,}\nonumber \\
D^{\rm NGL}_{\rm c}(B^Z){\rm d}B^Z\propto-2\pi r{\rm d}r& = 2\pi {1 \over -B''}{\rm d}B^Z & {\rm for~NGL.} \nonumber
\end{align}
Note that above we consider that a positive d$B^Z$ corresponds to a negative ${\rm d}r$.
Consequently the ratio of probability field distributions is
${\mathcal R} = D^{\rm NGL}_{\rm c}(B^Z)/D^{\rm BCS}_{\rm c}(B^Z) = B'^2/(-\delta B^Z B'')$.
For an intervortex distance $a$ the first
derivative of the conically shaped vortex lattice $B' = -2 B_{\rm pp}/a$, while for the
cosine second derivative $B'' = -4\pi^2 B_{\rm pp}/a^2$.
Here $B_{\rm pp}$ denotes difference between maximal and minimal fields along the line between
the neighboring vortices. Therefore the ratio of spectral weights of probability field distribution
in the vicinity of vortex core can be expressed as: ${\mathcal R} =  B_{\rm pp}/\pi^2\delta B^Z$.
Since $\delta B^Z\ll B_{\rm pp}$ close to the vortex core the spectral weight of the NGL model
is appreciably larger than that of the BCS model. Considering next terms in the Taylor expansion
this is valid for $\delta B^Z \lesssim B_{\rm pp}/4$.
The region of the field distribution close to $B^Z_{\rm vc}$ is a direct
fingerprint of the spatial field profile close to the vortex core.
Reduction of $D_{\rm c}^{\rm exp}(B^Z)$ spectral weight
in the field range marked with $\delta B'$ in  Fig.~\ref{fig:DcBdiff}
directly points to the narrowing of vortex core towards  the conical-like shape.
As can be seen a smearing does not reduce appreciably this spectral weight.
Experimentally the smearing strength can be estimated from the sharpness of the
cut-off in $D_{\rm c}^{\rm exp}(B^Z)$
near $B^Z_{\rm vc}$ (indicated by the arrow in Fig. \ref{fig:DcBdiff}).
Thus, the reduction of the spectral weight in the experimental spectrum
is basically determined by the field profile in the vicinity of vortex core.

\subsection{Vortex lattice pinning}

\begin{figure}
\includegraphics[width=0.9\linewidth,trim=0.6cm 0.06cm 0.8cm 0.8cm, clip=true]{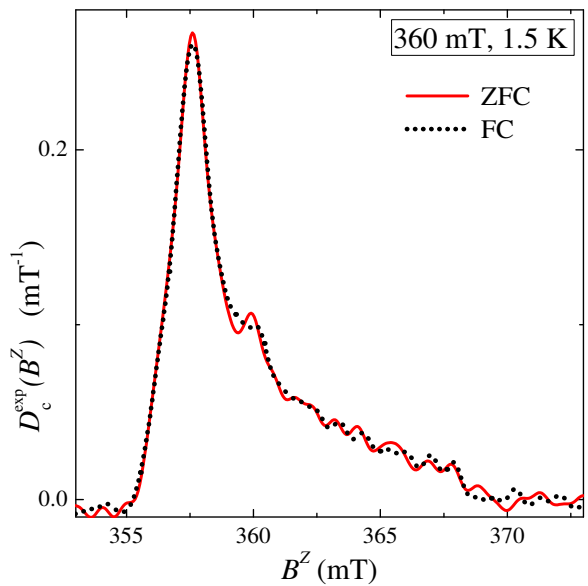}
\caption{(Color online) Additional test of the vortex pinning effect on $D^{\rm exp}_c(B^Z)$
carried on the Dolly instrument at  PSI. Field cooled (FC) and zero-field cooled (ZFC) experiments
were performed at 1.5 K and 360 mT. These field distributions coincide within precision of
the experiments which confirms the high purity of the Nb sample and
its low concentration of pinning centers. }
\label{fig:ZFCuFC}
\end{figure}

It is known that a vortex lattice may be pinned on local defects of the crystal lattice.
The pinning strength substantially depends on concentration and character of pinning centers.
For dirty Nb superconducting samples with a large concentration of pinning centers and
a large upper critical field ($B_{\rm c2}\gtrsim 600$ mT) small angle neutron scattering (SANS)
experiments showed a Bragg glass state with a diffraction pattern which depends on the history of applied
field and temperature.\cite{Daniilidis07} Enhanced pinning of the vortex lattice may
occur on the surface due to usually larger defect concentration at the surface
of a sample.\cite{Hanson11} These effects may be even more pronounced in the vicinity
of $B_{\rm c2}$ due to the peak effect. In order to exclude vortex lattice distortions due to
pinning and measure a sample in a condition close to thermodynamic equilibrium, a field
cooling procedure is usually used. Measurements are performed after relaxation of the vortex lattice
to its equilibrium state.
Extreme deviation of VL from its equilibrium state may be obtained in zero field cooling (ZFC) condition.
In this case the sample is cooled in zero field and afterwards a field is applied. Vortices cross the
surface and travel across the sample (see e.g. Ref.~\onlinecite{Qviller12}). If there are even moderate
concentration of pinning centers the VL will be far from its equilibrium state (see e.g. Ref.~\onlinecite{Sonier00}).
In Fig. \ref{fig:ZFCuFC} we show ZFC and FC measurements of $D_c^{\rm exp}(B^Z)$ for our Nb sample
at the lowest temperature (1.5 K) and 360 mT carried out on the Dolly instrument at the Paul Scherrer Institute.
Note that the pinning is expected to be strongest at the lowest temperatures due to
the reduction of the thermal excitations of VL. Within experimental uncertainties these
two field distributions coincide
confirming the high purity of the sample and its low concentration of pinning
centers. This experiment demonstrates that deviations from Ginzburg-Landau predictions summarized in the
main text cannot be ascribed to pinning effects in the sample.

\subsection{Some details of analysis}

As described in the main text in the clean limit a form factors and consequently a field
distribution depends on the three parameters $\tilde{a}$, $\tilde{b}$, and $\tilde{c}$.
All of them are temperature and field dependent which are determined by the temperature
and field dependence of $\Delta_0$ and temperature dependence of $B_{\rm c2}$.
According to Eq. (36) of Ref.~\onlinecite{MaisuradzeYaouanc13arXiv}
$\tilde{b}$ can be expressed as follows:
\begin{equation}
\tilde{b} = \frac{1}{\pi^2} \frac{\xi_{\rm GL}^2(T)}{\xi_0^2(T)}\frac{1-b}{b}  \simeq 0.110\frac{1-b}{b}.
\label{eq:TildeB}
\end{equation}
Above, $\xi_{\rm GL}(T)=\sqrt{\Phi_0/2\pi B_{\rm c2}(T)}$, $\xi_0(T)=\hbar v_{\rm F}/\pi \Delta_0(T)$,
where $\xi_{\rm GL}(0)/\xi_0(0) \simeq 1.04$ is the ratio of the Ginzburg-Landau and BCS coherence
lengths (which is assumed to be temperature independent)
\cite{MaisuradzeYaouanc13arXiv} and  $b = \overline{B^Z}/B_{\rm c2}(T)\simeq B_{\rm ext}/B_{\rm c2}(T)$.
Note that according to Eq.~(E2) of Ref.~\onlinecite{MaisuradzeYaouanc13arXiv}
for $3\tilde{b}\ll 1$ there is only a weak dependence of the form factors on $\tilde{b}$.
Temperature and field dependence of $\tilde{c}$ are defined as follows:
\begin{equation}
\tilde{c} = \frac{\Lambda}{\xi^T}.
\label{eq:TildeC}
\end{equation}
Using the definitions $\Lambda = \sqrt{\Phi_0/(2\pi \overline{B^Z})}$ and
$\xi^T = \hbar v_{\rm F}/(2\pi k_{\rm B}T)$ we obtain Eq. (4) of the main text.
The parameter $\tilde{a} = -\mu_0 N_0 \Delta^2_0 {\tilde c}/2{\overline {B^Z}}$
is related to the temperature and field dependence of $\Delta^2_0$.
In table \ref{table1} we provide a summary of the fit parameters for all the
data in the main text using the BCS-Gor'kov model.
\begin{table}[!tb]
\caption[~]{Summary of the fit parameters using the BCS-Gor'kov model [Eqs. (1)-(3) of the main text]
for the data presented in the main text. Note that the parameter $\tilde{b}$
was fixed according to Eq.~(\ref{eq:TildeB}) while $\tilde{c}$ was determined
with Eq.~(\ref{eq:TildeC}) (e.g. leaving common $v_{\rm F}$ free for all the
spectra).  The error in the temperature includes the magnitude
of the temperature fluctuations.  }\label{table1}
\begin{center}
\begin{tabular}{l l l l l l l }
\hline
\hline
$T$  &$B_{\rm ext}$    &$\tilde{a}$   & $\tilde{b}$  & $\tilde{c}$  & $\tilde{d}$ & $\sigma_{\rm s}$ \\
 (K)    &(mT)     &(mT)                   &     &   &     & (mT) \\
\hline
0.020(1)~~&360~~  &$0.623(1)$~~    & 0.021~~        &      0.08~~~     &    0.01~~  &  0.44(1) \\
0.100(1)&360   &$0.623(3)$     & 0.021        &      0.08     &    0.01  &  0.57(2)    \\
0.500(1)&360   &0.599(4)       & 0.021        &      0.08     &    0.01  &  0.43(1)   \\
0.800(1)&360   &0.770(5)       & 0.019        &      0.105(11)    &    0.01  &  0.39(1)     \\
1.200(1)&360   &1.135(8)       & 0.017        &      0.157(16)    &    0.01  &  0.38(1)      \\
1.600(1)&360   &1.328(8)       & 0.014        &      0.201(20)    &    0.01  &  0.34(1)      \\
3.00(15)&300    &4.56(3)       & 0.018        &      0.396(40)    &    0.01  &  0.49(1)      \\
4.50(2)&240    &7.07(5)       & 0.012        &      0.590(60)    &    0.01  &  0.48(1)     \\
5.90(2)&165    &12.6(1)       & 0.011        &      0.776(80)    &    0.01  &  0.58(1)       \\
7.80(2)&54     &115(1)      & 0.038        &      1.75(18)     &    0.01  &  0.60(1)        \\
\hline
\hline
\end{tabular}
\end{center}
\end{table}

\subsection{Conclusions}

To conclude, from zero-field $\mu$SR measurement we estimate the magnitude of
the muon diffusion rate $D_{\mu}$ at 1.6 K in the
studied Nb sample (i.e. $D_{\mu} < 0.01$ nm$^2\mu$s$^{-1}$). This value is found to be well below
the sensitivity limit of the TF $\mu$SR experiments ($D_{\mu} =1$ nm$^2\mu$s$^{-1}$, see Ref.~\onlinecite{Seeger79})
which we demonstrate by presenting the simulated TF field distributions.
These simulations exclude any appreciable influence of such a low diffusion rate on the measured field
distribution $D_c^{\rm exp}(B^Z)$. Measurements of $D_c^{\rm exp}(B^Z)$ in zero field cooled and field cooled
conditions at 1.5 K and 360 mT show that the vortex lattice pinning is very small and cannot influence the
deviations from the Ginzburg-Landau predictions presented in the main text.
In addition, details of data analysis and further results are presented.


\end{document}